\documentclass{PoS}

\title{An updated view on the parent population of\\ 
$\gamma$-ray emitting narrow-line Seyfert 1 galaxies}

\ShortTitle{Parent population of radio-loud NLS1s}

\author{\speaker{M. Berton}\\
	Dipartimento di Fisica e Astronomia "G. Galilei", Universit\`a degli Studi di Padova, Vicolo dell'Osservatorio 3, 35122, Padova (Italy)\\
	INAF - Osservatorio Astronomico di Brera, Via E. Bianchi 46, 23807, Merate (Italy) \\
        E-mail: \email{marco.berton@unipd.it}}

\abstract{The discovery of $\gamma$-ray emission from some radio-loud narrow-line Seyfert 1 galaxies (RLNLS1s) indicated the presence of a relativistic beamed jet in these objects. This immediately opened a new question about their parent population: is how do they look like when observed at large angles? The study of the parent population has provided important new insights on the physical properties of this fascinating class of active galactic nuclei (AGN). In this work, I will review the most recent advances in the study of parent sources, and I will show the impact that a recent continuum survey carried out with the Karl G. Jansky Very Large Array had on this topic. Finally, I will show how all these new results point toward a new unification model for young jetted AGN. }

\FullConference{Revisiting narrow-line Seyfert 1 galaxies and their place in the Universe - NLS1 Padova\\
		9-13 April 2018 \\
		Padova Botanical Garden, Italy}

\begin{document}

\section{Introduction}
Since their first classification \cite{Osterbrock85}, the nature of narrow-line Seyfert 1 galaxies (NLS1s) has been strongly debated. This special class of active galactic nuclei is characterized by the full width at half maximum (FWHM) of H$\beta$ $<$2000 km s$^{-1}$, the flux ratio [O III]$\lambda 5007$/H$\beta$ $<$3, and strong Fe II emission, usually (e.g. see \cite{Rakshit17}). The narrowness of permitted lines can be interpreted in two ways. The first one is a low rotational velocity around a low-mass black hole accreting close to the Eddington limit, a sign of the young age of the AGN \cite{Mathur00}. The second is a pole-on view of a disk-like broad-line region (BLR), which produces narrow lines because of low projected rotational velocity \cite{Decarli08}. The low-inclination scenario became more popular after the discovery of $\gamma$-ray emission coming from some radio-loud\footnote{Radio-loudness is defined as the ratio between the optical flux density in B-band and the radio flux density at 5 GHz \cite{Kellermann89}. Objects with ratio $>$10 are radio-loud. This criterion is highly debated \cite{Padovani17, Lahteenmaki17}.} NLS1s, indicating the presence of a relativistic beamed jet in these objects \cite{Abdo09}. Jetted-NLS1s represent a small fraction of the whole NLS1 popoulation, not far from 5\% \cite{Cracco16, Rakshit17}, but their existence challenged the well-known paradigm in which only high-mass black holes in giant elliptical galaxies are able to launch relativistic jets \cite{Laor00}. This led some authors to support the low-inclination hypothesis, in order to reconcile the new observations with the old paradigm (e.g. \cite{Calderone13, Dammando14, Baldi16, Miller17}). However, much observational evidence clearly indicated that this conclusion cannot be valid for the vast majority of jetted-NLS1s. Therefore, they represent a third new class of $\gamma$-ray emitting AGN \cite{Abdo09b}. In the following, I will show how the search for the parent population of $\gamma$-NLS1s provided new fundamental pieces of information about the nature of NLS1s, all pointing in the direction of a young evolutionary phase of AGN.  

\section{The parent population problem}
To date, the known number of $\gamma$-ray emitting (beamed) NLS1s is 19 \cite{Foschini15, Dammando15, Liao15, Berton17, Paliya18}, and all but one \cite{Berton17} are included in the larger class of flat-spectrum radio-loud NLS1s (F-NLS1s). Assuming a bulk Lorentz factor of $\sim$10, which is reasonable for $\gamma$-NLS1s \cite{Abdo09b}, and zero opening angle of the jet, we would expect $\sim19\times2\Gamma^2 = 3800$ parent population (unbeamed) objects. Even if a conservative conical opening angle of the torus of 30$^\circ$ is assumed, then $\sim$2/3 of these objects are type 2 AGN, and the number of parent NLS1s should yet be almost twice that of currently known radio-loud NLS1s ($\sim$600, \cite{Rakshit17, Chen18}). This lack of misaligned counterparts is called the parent population problem \cite{Foschini11}. \par
As parent population sources, three groups of candidate have been proposed. The first was steep-spectrum radio-loud NLS1s (S-NLS1), that is NLS1s with a relativistic jet observed at large angles. Such objects are the most obvious misaligned counterparts, but they are even rarer than F-NLS1, so they do not solve the parent population problem. A second hypothesis, made to take into account both the low-inclination model and the type 2 parent population, is that of disk-hosted radio galaxies. A flat BLR observed at high inclination would make the permitted lines broader (and then narrow because of torus obscuration in type 2 objects). Given that NLS1s are typically hosted by disk galaxies \cite{Crenshaw03}, a parent source would then appear as a broad- or narrow-line radio galaxy with a disk host. Even in this case, the number of known sources is relatively low (e.g., \cite{Mao14}), so they represent a small improvement to the parent population problem. The third and last option is that the parent population is hidden among radio-quiet (or radio-silent, that is without a radio detection) sources. A flat-spectrum object with an observed flux of 50 mJy, for example, would have an observed flux of 0.5 mJy if seen at large angles, assuming typical jet parameters \cite{Berton18}. Such a low flux is not detectable by the FIRST survey \cite{Becker95}, therefore the source would be classified as radio-quiet/silent. \par

\section{The contribution of optical spectroscopy}
Single-epoch spectroscopy has been fundamental in the analysis of this problem, especially by means of providing estimates of black hole mass and Eddington ratio. The black hole mass has often been estimated using the FWHM(H$\beta$) as a proxy for the gas velocity. However, many authors claimed that the FWHM can be severely affected by the BLR geometry, thus providing misleading results \cite{Decarli08}. Another way to derive the velocity is by means of the second-order moment $\sigma$ of H$\beta$. The latter provides better results than FWHM when used for low-contrast lines, and also has a lower uncertainty \cite{Peterson11}. \par
The mass of F-NLS1s \cite{Foschini15} and of their putative parent sources \cite{Berton15} were then derived by using $\sigma$. This different approached changed only marginally the black hole mass estimates with respect to previous estimates. The distributions of black hole mass and Eddington ratio suggest that the best candidates as parent sources are S-NLS1s and, partially, disk-hosted radio galaxies. Radio-quiet NLS1s instead seem to have systematically smaller black hole mass. Therefore, they seem to be a different class of objects with respect to radio-loud sources. \par
The same result has been suggested by the analysis of [O III]$\lambda\lambda$4959,5007 line profiles \cite{Berton16a}. These lines in NLS1s present peculiar properties. Typically each line profile can be reproduced by a core component, which is associated with gas of the narrow-line region (NLR), and a wing component, associated with an outflow coming from the central engine, and systematically blue-shifted (approaching the observer). In some NLS1s, the [O III] lines are shifted toward shorter wavelengths with respect to low-ionization lines, indicating bulk motion in the NLR \cite{Marziani03, Komossa08}. Sources showing this feature are called blue outliers, but this property is not present with the same frequency in all NLS1s. When investigated separately, radio-quiet and radio-loud sources behave differently. In particular, blue outliers are significantly more frequent among radio-loud objects (16/54 vs 4/68), possibly because they are attributed to the interaction between the relativistic jet and the interstellar medium. This result also suggests that radio-quiet objects simply lack relativistic jets, and therefore are not part of the parent population. \par
The problem, at this point, is still open. However, the conclusions that were derived from optical spectroscopy might have a fundamental flaw. All of the sources analyzed in the optical have a radio counterpart in the FIRST survey. This means that objects as faint as that described above could not be included in the study. This likely creates a selection effect, that unfortunately could not be avoided. At least in this early stage of parent population search, in fact, it was reasonable to include only radio-detected objects to identify misaligned radio sources. To avoid this problem, the search continued in the radio domain, where the lack of data represents a serious challenge to any analysis. 

\section{The role of compact steep-spectrum objects} 
An interesting class of radio sources, widely believed to be young radio galaxies, is that of compact steep-spectrum objects (CSS, \cite{Odea98}). Several authors have pointed out a possible connection between these objects and NLS1s (e.g., \cite{Oshlack01}), often noting that the radio morphology of CSS and, in particular, S-NLS1s, are very similar \cite{Gu15, Caccianiga17}. The hypothesis that CSS constitute to the parent population was investigated by means of the radio luminosity function (LF) at 1.4 GHz \cite{Berton16b}. The LF is particularly helpful when comparing a beamed population with an unbeamed one, because it is possible to analytically add a model of relativistic beaming to the unbeamed LF, and later compare the result with the observed LF of the beamed population \cite{Padovani92}. Unlike other comparisons between NLS1s and CSS, this required statistically complete samples. Given the low number of known F-NLS1s, the samples were inevitably small, 13 F-NLS1s and 12 CSS all classified as high-excitation radio galaxies (HERG), plus a control sample of 50 flat-spectrum radio quasars (FSRQs). It is worth noting that this study includes only the brightest objects, and that it was devoted only to HERG sources. NLS1s and HERGs indeed are both characterized by an efficient accretion mechanism \cite{Heckman14}. The first result of this study is that the LF of F-NLS1s is the low-luminosity tail of the FSRQs LF, showing again how jet power and radio luminosity are scalable (non-linearly) with the black hole mass \cite{Heinz03}. The second result is that CSS/HERGs, especially those with lower luminosities, can belong to the parent population of F-NLS1s. In particular, the beaming model reproduces satisfactorily the data when the ratio between the unbeamed (diffuse) emission and the beamed (jet) emission is close to 1. In regular blazars, this ratio is very small, because the diffuse emission is much larger than the jet emission (100 or 1000 times larger, \cite{Padovani92}). Therefore, if this unification is correct, we expect to observe weak or negligible diffuse emission in F-NLS1s. 

\section{The JVLA survey}
In order to have a better idea of the nature of the kpc-scale emission in NLS1s, a survey of 79 sources at 5 GHz has been carried out using the Karl G. Jansky Very Large Array (JVLA). This JVLA survey is the largest and deepest continuum survey of NLS1s carried out to date \cite{Berton18}. Thanks to the recent upgrade of the JVLA, the survey images reached a noise level of 10 $\mu$Jy in 10 minutes of exposure time per source. Out of the 79 sources, two were lost by pointing errors, two were later classified as intermediate-type Seyfert galaxies and excluded from the sample, and one, Mrk 739, was too faint to be detected. This last result is rather unexpected, since the source was reported to have a flux of 1.3$\pm$0.3 mJy at 5 GHz \cite{Koss11}. The reason for this discrepancy is not clear. \par
The main result of this survey was that radio-quiet and radio-loud NLS1s have typically different morphology. Radio-quiet objects tend to exhibit more spreaded-out emission, while radio-loud objects are typically more core-dominated. Some differences were observed also among flat- and steep-spectrum radio-loud NLS1s. While the latter have a somewhat hybrid morphology, with a relatively high fraction of diffuse emission (between 5 and 25\% of the total flux), the former show a very compact morphology, with only a few exceptions. Specifically, only three out of 28 objects have an extended morphology, while more than half (54\%) are core-dominated (diffuse flux below 5\%). \par
The upper limits on brightness temperature that could be measured in the JVLA images provided some insights on the origin of the radio emission. Very low brightness temperatures are observed in radio-quiet sources, pointing in the direction of strong ongoing star formation in these young objects \cite{Sani10}. Conversely, all radio-loud sources have higher temperatures, especially $\gamma$-ray emitters, confirming the non-thermal origin of the radio emission. \par

\section{The strange case of Mrk 783}
Another interesting result of the JVLA survey is the first detection of relic emission in a NLS1 \cite{Congiu17a}. The object is Mrk 783, a well-known source already classified in the original paper by Osterbrock \& Pogge \cite{Osterbrock85}, which in terms of radio luminosity lies close to the divide between radio-quiet and radio-loud objects. At 5 GHz, its core shows a steep spectral index close to 0.7 (F$_\nu \propto \nu^{-\alpha}$), which is consistent with synchrotron emission. On kpc scale the spectrum becomes however significantly steeper, $\sim$2. This extreme value is not common in AGN, and can be interpreted as the sign of a possible relic nature of the diffuse emission. If this hypothesis is correct, Mrk 783 would be one of the few sources in which the duty cycle of an AGN is observed in action. The diffuse emission might have been produced during past activity phase of the nucleus. Without the constant injection of energy from the AGN, the high-energy electron population of the relic becomes depleted, producing with time a steeper and steeper spectral index that we observe. \par
Finally, another remarkable feature observed in Mrk 783 is the presence of an extended narrow-line region 35 kpc away from the AGN (projected distance, \cite{Congiu17b}). This structure is the largest of its kind known to date at low redshift, the previous one being NGC 5252 ($\sim$25 kpc, \cite{Tadhunter89}). The emission is aligned with the radio axis, and it can likely be attributed to ionization by AGN. For this reason, Mrk 783 represents also an important example of relative interplay between an AGN and its host galaxy. 

\section{The unification of young jetted AGN}
The most important result obtained by the JVLA survey is that F-NLS1s have very dim diffuse emission. This is in agreement with the prediction of the radio LF. This behavior is completely different than that of classical blazars. At kpc scale, blazars do indeed show very bright unbeamed emission \cite{Antonucci85}, a sign that their jets are much more powerful. This could be an indication that F-NLS1 black hole mass is different compared to classical blazars, in agreement with the scaling of relativistic jets \cite{Heinz03}. The large-scale environment also supports this conclusion, since F-NLS1s reside in a different environment than FSRQs. Only 12\% of radio-loud NLS1s, but more than 50\% of FR II radio galaxies (associated with FSRQs), reside in superclusters \cite{Jarvela17}. The cluster environment, indeed, follows the galaxy mass, which in turn is connected to the black hole mass \cite{Ferrarese00}. On the other hand, it is also possible that, because of the young age, the radio lobes have not had enough time to fully develop. This result supports the unification scenario between radio-loud NLS1s and CSS --- especially those of low-luminosity --- as objects seen at different angles. \par 
The scenario depicted by this model is quite interesting. A possible evolutionary path can indeed be present among fast-accreting jetted AGN, with F-NLS1s and CSS being the beamed and unbeamed progenitors of FSRQs and FR$_{\rm HERG}$ radio galaxies, respectively. What triggers the radio jet is not yet clear, but merging is the usual suspect. Some radio-loud NLS1s show distinct signs of interaction with other galaxies \cite{Anton08, Jarvela18}. In this sense, radio-loud NLS1s might be different from their radio-quiet counterparts. Radio-quiet NLS1s are indeed believed to grow via secular evolution \cite{Mathur12}, while radio-loud might predominantly originate via merging (e.g., see \cite{Chiaberge15}). The different redshift distribution, in addition, could also play a role, since it is known that at high redshift more massive elliptical galaxies are more common than in the present-day Universe. \par
In conclusion, a tentative unification between radio-loud NLS1 galaxies and (low-luminosity) CSS objects is possible. If, as many believe, these objects are young AGN, this new model can describe their properties in terms of orientation. However, as mentioned above, this scenario is valid only for sources with efficient accretion mechanisms (i.e., HERGs). Whether an analog situation can be found among non-efficient accretors (i.e., LERGs) is still not clear, and should be investigated in the future to provide a complete unification of jetted AGN \cite{Foschini17b}.

\section*{Acknowledgements} 
This conference has been organized with the support of the Department of Physics and Astronomy ``Galileo Galilei'', the University of Padova, the National Institute of Astrophysics INAF, the Padova Planetarium, and the RadioNet consortium. RadioNet has received funding from the European Union's Horizon 2020 research and innovation programme under grant agreement No~730562. The author is thankful to A. Caccianiga, S. Ant\'on, and B. Peterson for helpful suggestions and discussion.

\end{document}